\documentstyle[12pt, axodraw]{article}
\newcommand{\bc} {\begin{center}}
\newcommand{\ec} {\end{center}}
\newcommand{\be} {\begin{equation}}
\newcommand{\ee} {\end{equation}}
\def\lsim{\mathrel{\rlap{\lower4pt\hbox{\hskip1pt$\sim$}}
    \raise1pt\hbox{$<$}}}         

\def\gsim{\mathrel{\rlap{\lower4pt\hbox{\hskip1pt$\sim$}}
    \raise1pt\hbox{$>$}}}         

\newcommand{\overstar}[1]{\mathrel{\rlap{\hbox{$#1$}}
    \raise3pt\hbox{$\hspace{0.5mm}^*$}}}
\makeatletter
\def\appendix{\par\clearpage
  \setcounter{section}{0}
  \setcounter{subsection}{0}
  \@addtoreset{equation}{section}
  \def\@sectname{Appendix~}
  \def\theequation{\thesection.\arabic{equation}}
  \def\thesection{\Alph{section}}}
\makeatother

\begin{document}

\begin{titlepage}
\hskip 11cm \vbox{ \hbox{Budker INP 2004-74}
\hbox{December 2004}} \vskip 0.3cm

\vskip 3.5cm \centerline{\bf  NON-FORWARD BFKL POMERON AT
NEXT-TO-LEADING ORDER$^{~\ast}$} \vskip 1cm \centerline{
V.S. Fadin$^{\dagger}$ and  R. Fiore$^{\ddagger}$ }

\vskip 1cm

\centerline{\sl $^{\dagger}$  Budker Institute for
Nuclear Physics and Novosibirsk State University,}
\centerline{\sl  630090 Novosibirsk, Russia}  \vskip .3cm
\centerline{\sl $^{\ddagger}$ Dipartimento di Fisica,
Universit\`a della Calabria} \centerline{\sl and Istituto
Nazionale di Fisica Nucleare, Gruppo collegato di
Cosenza,} \centerline{\sl I-87036 Arcavacata di Rende,
Cosenza, Italy}

\vskip 1cm
\begin{abstract}
The kernel of the BFKL equation for non-zero momentum
transfer is found at next-to-leading order. It is
presented  in various forms depending on the regularization
of the infrared singularities in ``virtual" and ``real"
parts of the kernel. The infrared safety of the total kernel
is demonstrated and a form free from the singularities is
suggested.

\end{abstract}
\vskip 1cm \vfill \hrule \vskip.3cm \noindent
$^{\ast}${\it Work supported in part by the Ministero Italiano
dell'Istruzione, dell'Universit\`a e della Ricerca, in
part by INTAS and
in part by the Russian Fund of Basic Researches.}\\
\vskip 0.5cm \vfill $\begin{array}{ll}
^{\dagger}\mbox{{\it email address:}} & \mbox{fadin@inp.nsk.su} \\
^{\ddagger}\mbox{{\it email address:}} & \mbox{fiore@cs.infn.it} \\
\end{array}$
\end{titlepage}
\eject

The kernel of the BFKL equation~\cite{BFKL} for the case
of forward scattering, i.e. for the momentum transfer
$t=0 $ and  vacuum quantum numbers in the $t-$channel,
was found at next-to-leading order (NLO) already five
years ago~\cite{FL98}. Unfortunately, the NLO calculation of
the kernel for  non-forward scattering was not completed
till now. We remind that the kernel depends on the
representation of the colour group in the $t$-channel; however
for any representation ${\cal R}$ it is given by the
sum of ``virtual" and ``real" contributions~\cite{FF98}.
The ``virtual" contribution is universal (does not depend
on ${\cal R}$). It is expressed through the NLO gluon
Regge trajectory \cite{trajectory} and is known.  The
``real" contribution is related to particle production in
Reggeon-Reggeon collisions and consists of parts coming
from one-gluon, two-gluon and quark-antiquark pair
production. The first part is expressed through the
effective Reggeon-Reggeon-gluon NLO vertex \cite{vertex}.
Apart from a colour  coefficient this part is  also
universal. It was found in Refs.~\cite{quark part of the
kernel} and \cite{gluon octet kernel} for the quark and
gluon contributions respectively.  Each of last two parts
for any ${\cal R}$ can be presented as a linear
combination of two independent pieces, one of which can
be determined by the antisymmetric colour octet
representation ${\cal R}=8_a$ (we shall call it gluon
channel) and the other by the colour singlet representation
${\cal R}=1$ (Pomeron channel). For the case of
quark-antiquark production both these pieces are known
\cite{quark part of the kernel}. Instead, only the piece
related to the  gluon channel is known for the case of
two-gluon production \cite{gluon octet kernel}.

The only missing piece of  the non-forward kernel was then
the two-gluon production contribution in the Pomeron
channel. We have calculated this contribution and
therefore have solved the problem of finding the
non-forward kernel at NLO for an arbitrary colour state in
the $t$-channel. Details of the calculation will be given
elsewhere. Here we present the NLO kernel for the most
important Pomeron channel. Note that for the case of the scattering of
physical (colourless) particles only the Pomeron channel
exists.  Since the quark contribution to the non-forward
kernel is known ~\cite{quark part of the kernel} for any
${\cal R}$, we shall consider in the following only the
gluon contribution, i.e.  pure gluodynamics.

Making use of the conventional dimensional regularization with
the space-time dimension $D=4+2\epsilon$, the BFKL
equation for the Mellin transform of the Green's
function of two Reggeized gluons in the $t$-channel is
written as
\begin{equation}\label{Green function}
\omega G \left( \vec q_1,\vec q_2;\vec q\,\right) = \vec
q_1^{\,2}\vec q_1^{\,\prime\,2}\delta ^{\left( D-2\right)
}\left( \vec q_1-\vec q_2\right) +
\int\frac{d^{D-2}r}{\vec r^{\:2}(\vec r-\vec q)^2}{\cal
K}\left( \vec q_1,\vec r;\vec q\right)G\left( \vec r,\vec
q_2;\vec q\,\right) ~,
\end{equation}
where  $q_i$ and $q_i^\prime \equiv q_i-q\,, \; (i=1\div
2)$ are the Reggeon (Reggeized gluon)  momenta, $q\simeq
q_\perp$ is the total $t$-channel momentum, $q^2\simeq
q^2_\perp =-\vec q^{~2}=t$ and the vector sign is used
for denoting the components of momenta  transverse to the
plane of initial momenta.  The kernel
\begin{equation}
{\cal K}\left( \vec q_1,\vec q_2;\vec q\,\right) = \left[
\omega \left( -\vec q_1^{\,2}\right) +\omega \left( -
\vec q_1^{\,\prime\,2} \right)\right] \vec q_1^{\,2}\vec
q_1^{\,\prime\,2}\delta ^{\left( D-2\right) }\left( \vec
q_1-\vec q_2\right) + {\cal K}_r\left( \vec q_1,\vec
q_2;\vec q\right) ~ \label{kernel=virtual+real}
\end{equation}
is given by the sum of the ``virtual" part, determined by
the gluon Regge trajectory $\,\omega(t)$ (actually the
trajectory $j(t)=1+\omega(t))$,  and the ``real" part,
related to particle production in Reggeon-Reggeon
collisions. In the limit $\epsilon \rightarrow 0$ we have
\cite{trajectory}
\[
\omega(t) =-2\bar
g_\mu^2\left(\frac{1}{\epsilon}+\ln\left(\frac{-t}{\mu^2}\right)\right)
-\bar g_\mu^4
\left[\frac{11}{3}\left(\frac{1}{\epsilon^2}-\ln^2\left(\frac{-t}
{\mu^2}\right)\right)+\left(\frac{67}{9}-2\zeta(2)\right)\right.
\]
\begin{equation}
\left.\times
\left(\frac{1}{\epsilon}+2\ln\left(\frac{-t}{\mu^2}\right)\right)
-\frac{404}{27} +2\zeta(3) \right]~. \label{renormalized
NLO trajectory}
\end{equation}
Here
\begin{equation} \bar g_\mu
^2=\frac{g_\mu ^2N_c\Gamma (1-\epsilon )}{(4\pi
)^{2+{\epsilon }}}~,  \label{coupling renormalization}
\end{equation}
$g_\mu $ being the renormalized coupling in the
${\overline{MS}}$ scheme, $N_c$ is the number of colors,
$\Gamma(x)$ is the Euler function and $\zeta(n)$ is the
Riemann zeta function, ($\zeta(2)=\pi^2/6$).

The remarkable properties of the ``real" part of the
kernel, which follow from general arguments, are
\begin{equation}
{\cal K}_{r}( 0,\vec{q}_{2};\vec{q}\,) = {\cal K}_{r}(
\vec{q}_{1},0;\vec{q}\,) ={\cal K}_{r}(
\vec{q},\vec{q}_{2};\vec{q} \,) ={\cal K}_{r}(
\vec{q}_{1},\vec{q};\vec{q}\,) =0
 \label{gauge invariance of the kernel}
\end{equation}
and
\begin{equation}
{\cal K}_{r}( \vec{q}_{1},\vec{q}_{2};\vec{q} \,)= {\cal
K}_{r}( -\vec{q} _{1}^{\, \prime},-\vec{q}_{2}^{\, \prime
};\vec{q}\,) ={\cal K}_{r}( -\vec{q}_{2},-\vec{q}_{1};-
\vec{q}\,) ~. \label{symmetry of the kernel}
\end{equation}
The properties (\ref{gauge invariance of the kernel})
imply that the kernel turns into zero at zero transverse
momenta of the Reggeons and appear as consequences of the
gauge invariance; in turn the properties (\ref{symmetry of
the kernel}) are the consequence of cross-invariance.

In pure gluodynamics the ``real"  part ${\cal K}_r$ is
given by sum of one-gluon-  and two-gluon-production
contributions. The first of them differs from the
corresponding contribution in the gluon channel only by a
colour group coefficient. As for the second one, it occurs to be
much more complicated  in the Pomeron channel than in the gluon
one. The simplicity of the gluon channel is related to the
gluon Reggeization. Technically it is determined by the
cancellation of contributions of non-planar diagrams due
to the colour group algebra. The complexity of contributions of
non-planar diagrams is well known  since the calculation
of the non-forward kernel for the QED Pomeron
\cite{Non-forward Pomeron in QED} which was found only in
the form of a two-dimensional integral. In QCD the situation
is greatly worse because of the existence of cross-pentagon
and cross-hexagon diagrams in addition to QED-type
cross-box diagrams.  It requires the use of additional
Feynman parameters. At arbitrary $D$ no integration over
these parameters at all can  be done in elementary
functions. It occurs, however, that in the limit
$\epsilon \rightarrow 0$ the integration over additional
Feynman parameters can be performed, so that the result
can be written as a two-dimensional integral, as well
as in QED.

Let us present the kernel ${\cal K}_r$ in the limit
$D=4+2\epsilon \rightarrow 4$ as sum of two parts:
\begin{equation}
{\cal K}_r={\cal K}^{sing}_r + {\cal K}^{(reg)}_{r}~.
\label{K1 through Ksing and Kreg}
\end{equation}
Here the first contains all singularities:
\[
{\cal K}_r^{sing}(\vec q_1,\vec q_2; \vec{q})
=\frac{2\bar g_\mu ^2\mu^{-2\epsilon}}{\pi
^{1+{\epsilon}}\Gamma(1-\epsilon)}\left(
\frac{\vec{q}_1^{\:2} \vec{q}_2^{\:\prime\:2}+
\vec{q}_1^{\:\prime\:2} \vec{q}_2^{\:2}}{\vec k ^{\:2}}-
\vec{q}^{\:2}\right)\Biggl\{ 1+ \bar g_\mu
^2\biggl[\frac{11}{3\epsilon}
\]
\begin{equation}
+\left(\frac{\vec k^{\:2}}{\mu^2}\right)^\epsilon
\biggl\{-\frac{11}{3\epsilon}+\frac{67}{9}
-2\zeta(2)+\epsilon\left( -\frac{404}{27} +14\zeta(3)
+\frac{11}{3}\zeta(2)\right)\biggl\}\biggr]\Biggr\}~,\label{Ksingular}
\end{equation}
where $\vec k=\vec q_1-\vec
q_2=\vec{q}_1^{\:\prime}-\vec{q}_2^{\:\prime}$.
The second, putting $\epsilon=0$ and $\bar
g_\mu^2=\alpha_s(\mu^2)N_c/(4\pi)$~, is given by
\[
{\cal K}_r^{reg}(\vec q_1,\vec q_2; \vec{q})
=\frac{\alpha_s^2(\mu^2)N^2_c}{16\pi^3} \Biggl[2(\vec
q_1^{\:2}+\vec q_2^{\:2}-\vec q^{\:2})
\left(\zeta(2)-\frac{50}{9}\right)
-\frac{11}{3}\left(\vec q_1^{\:2}\ln\left(\frac{\vec
q_1^{\:2}}{\vec k^{\:2}}\right)\right.
\]
\[
\left. +\vec q_2^{\:2}\ln\left(\frac{\vec q_2^{\:2}}{\vec
k^{\:2}}\right)-\vec q^{\:2}\ln\left(\frac{\vec
q_1^{\:2}\vec q_2^{\:2}}{\vec k^{\:4}}\right)-\frac{\vec
q_1^{\:2}\vec q_2^{\:\prime\:2}-\vec q_2^{\:2}\vec
q_1^{\:\prime\:2}}{\vec k^{\:2}}\ln\left(\frac{\vec
q_1^{\:2}}{\vec q_2^{\:2}}\right)\right)+\vec
q^{\,2}\left( \ln\left(\frac{\vec q_1^{\:2}}{\vec
q^{\:2}}\right)\ln\left(\frac{\vec q_1^{\:\prime 2}}{\vec
q^{\:2}}\right)\right.
\]
\[
\left.+\ln\left(\frac{\vec q_2^{\:2}}{\vec
q^{\:2}}\right)\ln\left(\frac{\vec q_2^{\:\prime 2}}{\vec
q^{\:2}}\right)+\frac{1}{2}\ln^2\left(\frac{\vec
q_1^{\:2}}{\vec
q_2^{\:2}}\right)\right)+\ln\left(\frac{\vec
q_1^{\:2}}{\vec q_2^{\:2}}\right)\left(\frac{\vec
q_1^{\:\prime 2}}{2}\ln\left(\frac{\vec q_2^{\:2}}{\vec
k^{\:2}}\right)-\frac{\vec q_2^{\:\prime
2}}{2}\ln\left(\frac{\vec q_1^{\:2}}{\vec
k^{\:2}}\right)\right.
\]
\[
\left.-\frac{\vec q_1^{\:2}\vec q_2^{\:\prime\:2}+\vec
q_2^{\:2}\vec q_1^{\:\prime\:2}}{2\vec
k^{\:2}}\ln\left(\frac{\vec q_1^{\:2}}{\vec
q_2^{\:2}}\right)+\frac{\vec q_1^{\:\prime\:2}(\vec
q_1^{\:2}-3\vec q_2^{\:2})}{2\vec
k^{\:2}}\ln\left(\frac{\vec k^{\:2}}{\vec
q_2^{\:2}}\right)+\frac{\vec q_2^{\:\prime\:2}(3\vec
q_1^{\:2}-\vec q_2^{\:2})}{2\vec
k^{\:2}}\ln\left(\frac{\vec k^{\:2}}{\vec
q_1^{\:2}}\right)\right)
\]
\[
+\left(\vec q^{\:2}(\vec k^{\:2}-\vec q_1^{\:2}-\vec
q_2^{\:2})+2\vec q_1^{\:2}\vec q_2^{\:2}-\frac{(\vec
q_1^{\:2}-\vec q_2^{\:2})(\vec q_1^{\:2}+\vec
q_2^{\:2})(\vec q_1^{\:\prime \:2}-\vec q_2^{\:\prime
\:2})}{2\vec k^{\:2}}+\vec q_1^{\:2}\vec q_1^{\:\prime
\:2}+\vec q_2^{\:2}\vec q_2^{\:\prime \:2}\right.
\]
\begin{equation}
\left.-\frac{\vec k^{\:2}}{2}(\vec q_1^{\:\prime
\:2}+\vec q_2^{\:\prime \:2})\right)I(\vec k^{\:2}, \vec
q_2^{\:2}, \vec q_1^{\:2})-2J(\vec q_1, \vec q_2; \vec
q)-2J(-\vec q_2, -\vec q_1; -\vec q)\Biggr ]
+\Biggl\{\vec q_i \longleftrightarrow -\vec
q_i^{\:\prime}\Biggr\}~. \label{Kregular}
\end{equation}
In expression~(\ref{Kregular}) two quantities appear, precisely
\begin{equation}
I(a, b, c)= \int_0^1\frac{dx}{ a(1-x)+b x-c x(1-x)}\ln
\left( \frac{ a(1-x)+ b x}{ c x(1-x)}\right)
\label{I(p,q,r)}
\end{equation}
and
\[
J(\vec q_1,\vec  q_2; \vec q)=\int_{0}^{1}dx\int_0^1
dz\left\{\vec q_1\vec q_1^{\:\prime}\left((2-x_1x_2)\ln
\left(\frac{Q^{2}}{\vec k^{\:2}}\right) -\frac{2}{x_1}\ln
\left(\frac{Q^{2}}{Q_0^{2}}\right)\right)\right.
\]
\[
\left.-\frac{1}{2Q^{2}} x_1x_2(\vec q^{\:2}_{1}-2\vec
q_{1}\vec p_{1})(\vec q^{{\:}\prime 2}_{1}-2\vec
q^{{\:}\prime~}_{1} \vec
p_{2})+\frac{2}{x_1}\left[\left(x_2\vec q_1\vec
q^{{\:}\prime~}_{1}(\vec p_1(\vec q^{{\:}\prime}_{1}
-\vec p_2))-\vec q^{{\:}\prime \:2}_{1}\vec q_1\vec
p_2\right)\frac{1}{Q^{2}}\right.\right.
\]
\[
\left. + \left(z(1-z)\vec q^{{\:}\prime \:2}_{2}\vec
q_1\vec q^{{\:}\prime~}_{1} +\vec q^{{\:}\prime\:
2}_{1}(z\vec q_1\vec k+(1-z)\vec q_1\vec
q^{{\:}\prime}_{1})\right)\frac{1}{Q_0^{2}}
\right]-\frac{1}{Q^2}\left(\vec q^{{\:}\prime \:
2}_{1}\vec q_1\left(\vec p_{1}-2\vec
q^{{\:}\prime}_{1}\right)\right.
\]
\[
\left.\left.  +4x_1\vec q_1^{\:2}(\vec
q^{{\:}\prime}_{1}\vec p_2)+\vec q^{{\:}\prime}_{1}\vec
q_{1}(\vec q^{{\:}\prime}_{1} \vec q_{1}-\vec
q^{{\:}\prime}_{1}\vec p_{1}-\vec q_{1}\vec p_{2})+2(\vec
q^{{\:}\prime}_{1}\vec p_{1}) (\vec q_{1}\vec p_2)-2(\vec
q^{{\:}\prime}_{1}\vec p_{2})(\vec q_{1}\vec p_1)\right)
\right.
\]
\[
\left. +{\vec q^{{\:}\prime
\:2}_{1}}\left[\frac{-1}{\mu_2^2Q^2}
\left(2\frac{x_2}{x_1}(\vec q_1\vec p_2)\vec
q^{{\:}\prime}_{1}\vec k+x_2(\vec q^{{\:}\prime}_{1} \vec
p_2)(\vec q_2^{\:2}-\vec k^{\:2})+2(\vec q_2\vec p_2)\vec
q_1\vec q\right)\right.\right.
\]
\[
+\frac{2}{\mu_0^2Q_0^2}\frac{1}{x_1}(\vec q_1\vec
p_0)\vec q^{{\:}\prime}_{1}\vec k -\frac{\vec q_1(\vec
q^{{\:}\prime}_{1}+\vec k)}{x_1}\left(\frac{x_2}{\vec
p_2^{\:2}}\ln
\left(\frac{Q^2}{\mu_2^2}\right)-\frac{1}{\vec
p_0^{\:2}}\ln \left(\frac{Q_0^2}{\mu_0^2}\right)\right)
\]
\[
+\frac{1}{\vec p_2^{\:2}}\left(\frac{1}{\vec
p_2^{\:2}}\ln\left(\frac{Q^2}{\mu_2^2}
\right)+\frac{1}{Q^2}\right) \left(2\frac{x_2}{x_1}(\vec
q_1\vec p_2)(\vec q^{{\:}\prime}_{1}+\vec k)\vec
p_2-2((x_2 \vec q^{{\:}\prime}_{1}+\vec q_2)\vec p_2)\vec
q_1\vec p_2\right)
\]
\[
-\frac{1}{\vec p_0^{\:2}}\left(\frac{1}{\vec
p_0^{\:2}}\ln\left(\frac{Q_0^2}{\mu_0^2}
\right)+\frac{1}{Q_0^2}\right) \left(2\frac{1}{x_1}(\vec
q_1\vec p_0)(\vec q^{{\:}\prime}_{1}+\vec k)\vec
p_0\right) +\frac{(x_2\vec q^{{\:}\prime}_{1}+\vec
q_2)\vec q_1}{\vec p_2^{\:2}}\ln\left(\frac{Q^2}
{\mu_2^2}\right)
\]
\[
+\frac{\vec q_1^{\:2}}{d}\left((\vec q_2\vec k)(\vec
q_2^{{\:}\prime}\vec k)\left(\frac{Q^2}{d}{\cal
L}-\frac{1}{\vec k^{\:2}} \right)+(\vec q_2\vec p_2)(\vec
q_2^{{\:}\prime}\vec k)\left(\frac{1}{\mu_2^2}
-\frac{\mu_1^2}{d}{\cal L}\right)+(\vec q_2\vec k)(\vec
q_2^{{\:}\prime}\vec p_1)\left(\frac{1}{\mu_1^2}
-\frac{\mu_2^2}{d}{\cal L}\right)\right.
\]
\begin{equation}
\left.\left.\left.+(\vec q_2\vec p_2)(\vec
q_2^{{\:}\prime}\vec p_1)\left(\frac{\vec
k^{\:2}}{d}{\cal L}-\frac{1}{Q^2}\right)+\frac{(\vec
q_2\vec q_2^{{\:}\prime})}{2}{\cal
L}\right)\right]\right\}~, \label{J3 at d=4}
\end{equation}
Here we make use of the following positions:
\[
\vec p_1=zx\vec q_1+(1-z)(x\vec k-(1-x)\vec
q_2^{\:\prime}),\;\;\vec p_2=z((1-x)\vec k-x\vec q_2)
+(1-z)(1-x)\vec q_1^{\:\prime}; \;\; \vec p_1+\vec
p_2=\vec k,
\]
\[
Q^2=x(1-x)(\vec q_1^{\:2} z+\vec q_1^{{\:}\prime
2}(1-z))+z(1-z)(\vec q_2^{\:2} x+\vec q_2^{{\:}\prime
2}(1-x)-\vec q^{\:2}x(1-x)),\;\;\;\mu_i^2=Q^2+\vec
p_i^{\:2},
\]
\[
\vec p_0= z\vec k +(1-z)\vec q_1^{\:\prime}; \;\;
Q_0^2=z(1-z)\vec q^{{\:}\prime \:2}_2,\;\;\;
\mu_0^2=z\vec k^{\:2}+(1-z)\vec q^{{\:}\prime\:
2}_1,\;\;\;
\]
\[
d=\mu_1^2\mu_2^2-\vec k^{\:2}Q^2=z(1-z)x(1-x)\left((\vec
k^{\:2}-\vec q_1^{\:2}-\vec q_2^{{\:}\prime {\:}2})(\vec
k^{\:2}-\vec q_1^{{\:}\prime {\:}2}-\vec q_2^{\:2})+\vec
k^{\:2}\vec q^{\:2}\right)+\vec q_1^{\:2}\vec
q_2^{\:2}xz(x+z-1)
\]
\begin{equation}\label{definition d and L}
+\vec q_1^{{\:}\prime {\:}2}\vec q_2^{{\:}\prime
{\:}2}(1-x)(1-z)(1-x-z),\;\;\; {\cal
L}=\ln\left(\frac{\mu_1^2\mu_2^2}{\vec
k^{\:2}Q^2}\right).
\end{equation}
Note that the integral $I(a,b,c)$ is invariant with
respect to any permutation of its arguments, which can be
seen from the representation
\begin{equation}
I(a,b,c)=\int_0^1\int_0^1\int_0^1\frac{dx_1 dx_2
dx_3\delta(1-x_1-x_2-x_3)}{(ax_1+ bx_2+ c
x_3)(x_1x_2+x_1x_3+x_2x_3)}~.\label{symmetric integral}
\end{equation}
In particular, $I(k^2, q_2^2, q_1^2)$ does not change performing
the substitution $q_1\leftrightarrow -q_2$.

As far as the quantity $J$ is concerned, its expression
(\ref{J3 at d=4}) is rather cumbersome. Unfortunately,
till now our attempts to find a more simple
representation for it have been unsuccessful.

All singularities of ${\cal K}_r$ are present only in its
first part ${\cal K}^{sing}_r$.  We recall that the one-gluon- and
two-gluon-production contributions to  ${\cal K}_r$
separately contain first and second order poles at
$\epsilon =0$. When summing these two contributions the pole
terms cancel, so that at fixed nonzero $\vec k^2$, when
the term $\left({\vec k^{\:2}}/{\mu^2}\right)^{\epsilon}$
in expression (\ref{Ksingular}) of the kernel
can be expanded in $\epsilon$, the
sum is finite at $\epsilon =0$.  However expression
(\ref{Ksingular}) is singular at $\vec k^{\:2}=0$ so
that, when it is integrated over $q_2$, the region of so
small $\vec k^{\:2}$, such that $\epsilon |\ln \left({\vec
k^{\:2}}/{\mu^2}\right)|\sim 1$, does contribute.
Therefore the expansion of $\left({\vec
k^{\:2}}/{\mu^2}\right)^{\epsilon}$ is not  done in expression
(\ref{Ksingular}). Moreover, the terms $\sim \epsilon$
are taken into account in the coefficient of the expression
divergent at $\vec k^{\:2}=0$, in order to save
all contributions non-vanishing in the limit $\epsilon\rightarrow 0$
after integration.

The part ${\cal K}_r^{({reg})}$  is finite in the limit
$\epsilon=0$. Moreover, integration of this part in
Eq.~(\ref{Green function}) for the Green's function
does not create singularities at $\epsilon=0$ as well.
Indeed, the points $\vec r=0$ and $\vec r^{\:\prime}=0$,
which  at first glance could give the singularities in
Eq.~(\ref{Green function}), are not dangerous
because of  the ``gauge invariance" properties (\ref{gauge
invariance of the kernel}) of the kernel ${\cal K}_{r}$.
It follows from formula (\ref{K1 through Ksing and Kreg}) that if
one of two parts (${\cal K}_r^{({sing})}$ or ${\cal
K}_r^{({reg})}$) of the kernel possesses these
properties, the same is valid for the other. ``Gauge
invariance" of ${\cal K}^{({sing})}_{r}$ is  evident from
expression (\ref{Ksingular}), therefore ${\cal K}_r^{({reg})}$ also
turns into zero at zero Reggeon momenta. It is worthwhile to
say that the fulfillment of these properties for ${\cal
K}_r^{({reg})}$ can be shown directly using the explicit
expression (\ref{Kregular}), although this is far from to be
evident.

As we have already seen, at $\epsilon = 0$ divergencies
can come from the region of small $\vec k$. Nevertheless,
it is not difficult to check from expression
(\ref{Kregular}) for ${\cal K}_r^{({reg})}$ that non-integrable
singularities at $\vec k=0$ are absent.

The total  kernel for the Pomeron channel must be
infrared safe. Infrared singularities of ${\cal K}_r$
must be cancelled by singularities of the gluon
trajectory after integration of the total kernel with any
function nonsingular at $\vec k =0$. Indeed, one can
easily see  that this is the case using
Eqs.~(\ref{renormalized NLO trajectory}) and
(\ref{Ksingular}).

It is convenient to present the total kernel in such a
form that the cancellation of singularities between real
and virtual contributions becomes evident. To this aim
let us first switch from the dimensional
regularization to the cut-off $\vec k^2 > \lambda^2$, with
$\lambda\rightarrow 0$,  which is more convenient for
practical purposes.  With such regularization we can pass
to the limit $\epsilon\rightarrow 0$ in the real part of
the kernel, so that for its  singular part we get
\[
{\cal K}_r^{sing}(\vec q_1,\vec q_2; \vec{q})\;
\rightarrow\; {\cal K}_{r}^{\lambda}(\vec q_1,\vec q_2;
\vec{q})=\frac{\alpha_s(\mu^2)N_c}{2\pi^2 }\left(
\frac{\vec{q}_1^{\:2} \vec{q}_2^{\:\prime\:2}+
\vec{q}_1^{\:\prime\:2} \vec{q}_2^{\:2}}{\vec k ^{\:2}}-
\vec{q}^{\:2}\right)
\]
\begin{equation}
\times \Biggl\{
1-\frac{\alpha_s(\mu)N_c}{4\pi}\left(\frac{11}{3}
\ln\left(\frac{\vec k^{\:2}}{\mu^2}\right)-\frac{67}{9}
+2\zeta(2)\right)\Biggr\}\theta((\vec q_1-\vec
q_2)^2-\lambda^2)~.
\end{equation}
The trajectory must be transformed in such a way that the cut-off
regularization yields the same result as the $\epsilon$
regularization does:
\[
\omega (t)\;\rightarrow \;\omega_{\lambda}(t)=
\lim_{\epsilon\rightarrow
0}\biggl(\omega(t)+\frac{1}{2}\int
\frac{d^{2+\epsilon}q_2}{\vec q_2^{\:2}\vec
q_2^{\:\prime\:2}}{\cal K}_r^{(1)}(\vec q_1,\vec q_2;
\vec q)\theta(\lambda^2-(\vec q_1-\vec q_2)^2)\biggr)
\]
\[
=-\frac{\alpha_s(\mu ^2)N_c}{2\pi}\biggl\{\ln
\left(\frac{-t}{\lambda^2}\right)-\frac{\alpha_s(\mu^2)N_c}{4\pi
}\left[
\frac{11}6\left(\ln^2\left(\frac{-t}{\mu^2}\right)\right.\right.
\]
\begin{equation}
\left.\left.-\ln^2\left(\frac{\lambda^2}{\mu^2}\right)\right)-\left(
\frac{67}9-\frac{\pi ^2}3\right)\ln
\left(\frac{-t}{\lambda^2}\right)+6\zeta
(3)\right]\biggr\}~.
\end{equation}
It is easy to check that by integrating over $d^2q_2$  any function
non-singular for $\vec k=0$ with the total kernel
(\ref{kernel=virtual+real}) at $\omega(t)\rightarrow
\omega_{\lambda}(t)$ and ${\cal K}_r^{sing}(\vec q_1,\vec
q_2; \vec{q})\; \rightarrow\; {\cal K}_{r}^{\lambda}(\vec
q_1,\vec q_2; \vec{q}) $  one obtains a $\lambda$
-independent result in the limit $\lambda\rightarrow 0$.
Moreover, it is also easy to find a form of the kernel which
does not contain $\lambda$ at all. It is sufficient to
find a representation
\begin{equation}\label{omega as integral}
\omega_{\lambda}(-\vec q_1^{\;2})=\int d^2q_2
f_\omega(\vec q_1,\vec q_2)\theta((\vec q_1-\vec
q_2)^2-\lambda^2)
\end{equation}
with  a function $f_\omega$ such that the non-integrable
singularities  at $\vec k=\vec q_1-\vec q_2=\vec
q_1^{\:\prime}-\vec q_2^{\:\prime}=0$ are  cancelled in
the ``regularized virtual kernel"
\begin{equation}
{\cal K}^{reg}_{v}(\vec q_1,\vec q_2;
\vec{q})=f_\omega(\vec q_1,\vec q_2)+f_\omega(\vec
q_1^{\:\prime},\vec q_2^{\:\prime})+\frac{{\cal
K}_{r}^{sing}(\vec q_1,\vec q_2;
\vec{q})|_{\epsilon=0}}{\vec q_2^{\:2}\vec
q_2^{\:\prime\:2}}~.
\end{equation}
After that we can proceed to the limit $\lambda=0$:
\[
\left(\hat{{\cal K}}\Psi\right)(\vec q_1)=\int d^2
q_2\biggl\{{\cal K}^{reg}_{v}(\vec q_1,\vec q_2;
\vec{q})\Psi(\vec q_1)
\]
\begin{equation}
+\frac{{\cal K}_{r}^{sing}(\vec q_1,\vec q_2;
\vec{q})|_{\epsilon=0}}{\vec q_2^{\:2}\vec
q_2^{\:\prime\:2}}\left(\Psi(\vec q_2)-\Psi(\vec
q_1)\right)+\frac{{\cal K}_{r}^{reg}(\vec q_1,\vec q_2;
\vec{q})}{\vec q_2^{\:2}\vec q_2^{\:\prime\:2}}\Psi(\vec
q_2)\biggr\}~.
\end{equation}
Of course, the choice of  the function $f_\omega$ contains
a large arbitrariness.  A simple choice is
\[
f_\omega(\vec q_1,\vec
q_2)=-\frac{\alpha_s(\mu^2)N_c}{2\pi^2 }
\frac{\vec{q_1}^{\:2}}{\vec k ^{\:2}(\vec{q_1}^{\:2}+\vec
k ^{\:2})}
\]
\begin{equation}
\times \Biggl\{
1-\frac{\alpha_s(\mu)N_c}{4\pi}\left(\frac{11}{3}
\ln\left(\frac{\vec k^{\:2}}{\mu^2}\right)-\frac{67}{9}
+2\zeta(2)\right)
+\left(6\zeta(3)-\frac{11}{3}\zeta(2)\right)\frac{\vec{k}^{\:2}}
{(\vec{q_1}^{\:2}+\vec k ^{\:2})} )\Biggr\}~.
\end{equation}

In a subsequent paper we shall present the results of the 
investigation of properties of the kernel.

 \vspace{0.2cm} \noindent \underline{\bf
Acknowledgments:} V.S.F. thanks the Dipartimento di
Fisica dell'Universit\`a della Calabria and the Istituto
Nazionale di Fisica Nucleare - gruppo collegato di
Cosenza for their warm hospitality while a part of this
work was done.

\end{document}